\documentclass[pra,twocolumn,superscriptaddress,showpacs]{revtex4}

\usepackage{amsmath}
\usepackage{amsfonts}
\usepackage{amssymb}
\usepackage{graphicx}

\newcommand{\half}{\tfrac{1}{2}}
\newcommand{\sq}[1]{\left[ {#1} \right]}
\newcommand{\ro}[1]{\left( {#1} \right)}
\newcommand{\Tr}{\operatorname{Tr}}
\newcommand{\ket}[1]{\ensuremath{\lvert #1 \rangle}}

\newcommand{\braket}[2]{\ensuremath{\langle #1|#2 \rangle}}
\newcommand{\ketbra}[2]{\ensuremath{\lvert #1 \rangle\langle #2 \rvert}}
\newcommand{\nn}{\nonumber \\}

\newcommand{\C}[1][{}]{\ensuremath{C^\text{#1}}}

\begin{document}
\title{Multiple-copy state discrimination: Thinking globally, acting locally}

\author{B.~L. Higgins}
\altaffiliation[Present address: ]{Institute for Quantum Computing, University of Waterloo, Waterloo, ON N2L 3G1, Canada}
\affiliation{Centre for Quantum Dynamics, Griffith University, Brisbane, 4111, Australia}
\author{A.~C. Doherty}
\affiliation{School of Physics, The University of Sydney, Sydney, 2006, Australia}
\author{S.~D. Bartlett}
\affiliation{School of Physics, The University of Sydney, Sydney, 2006, Australia}
\author{G.~J. Pryde}
\affiliation{Centre for Quantum Dynamics, Griffith University, Brisbane, 4111, Australia}
\author{H.~M. Wiseman}
\email{H.Wiseman@griffith.edu.au}
\affiliation{Centre for Quantum Dynamics, Griffith University, Brisbane, 4111, Australia}

\begin{abstract}
We theoretically investigate schemes to discriminate between two nonorthogonal quantum states given multiple copies. We consider a number of state discrimination schemes as applied to nonorthogonal, mixed states of a qubit. In particular, we examine the difference that local and global optimization of local measurements makes to the probability of obtaining an erroneous result, in the regime of finite numbers of copies $N$, and in the asymptotic limit as $N \rightarrow \infty$. Five schemes are considered: optimal collective measurements over all copies, locally optimal local measurements in a fixed single-qubit measurement basis, globally optimal fixed local measurements, locally optimal adaptive local measurements, and globally optimal adaptive local measurements. Here an adaptive measurement is one in which the measurement basis can depend on prior measurement results. For each of these measurement schemes we determine the probability of error (for finite $N$) and scaling of this error in the asymptotic limit. In the asymptotic limit, it is known analytically (and we verify numerically) that adaptive schemes have no advantage over the optimal fixed local scheme. Here we show moreover that, in this limit, the most naive scheme (locally optimal ﬁxed local measurements) is as good as any noncollective scheme except for states with less than 2\% mixture. For finite $N$, however, the most sophisticated local scheme (globally optimal adaptive local measurements) is better than any other noncollective scheme, for any degree of mixture.
\end{abstract}


\maketitle

\section{Introduction}

Understanding the principles governing the process of measurement has the potential to illuminate fundamental questions and practical notions of the behaviour of the physical world. Yet despite the exceptional experimental success of quantum theory, many questions regarding the fundamental principles of quantum measurement remain. A key aspect of this is the restriction imposed by quantum measurement that a measurement of an unknown quantum system cannot usually reveal complete information about that system. The consequences of this become evident in the problem of discriminating between two nonorthogonal quantum states~\cite{WisMil10}. Consider a quantum system that was prepared in one of two known states, but we do not know which one. The task is to determine which of the two preparations took place. If we consider only pure states, we may represent the two possibilities as $\ket{\psi_+}$ and $\ket{\psi_-}$. Except for orthogonal states (where $\lvert\braket{\psi_+}{\psi_-}\rvert = 0$) there is no measurement that could be applied that will deterministically find the state of the system without error~\cite{WisMil10}.

Despite this, it is possible to construct a measurement that always determines the correct state, but doing so produces a nonzero probability of obtaining an inconclusive result~\cite{WisMil10,Ivanovic1987,Dieks1988,Peres1988}. Alternatively, one can construct a measurement in which all results are conclusive, but necessarily possessing some probability of error, which is to be minimized~\cite{WisMil10,Helstrom1976}. This minimum-error-probability measurement is known as the Helstrom measurement, possessing an error probability $\C$ dependent upon the overlap of the states, $\lvert\braket{\psi_+}{\psi_-}\rvert$.

Should additional copies of the unknown state be available~\cite{Audenaert2007}, one can devise measurement schemes that exploit these additional copies to achieve a lowered error probability. An example is the simple ``majority vote'' scheme, where the Helstrom measurement is applied to each of $N$ copies individually, and the overall result chosen according to which state corresponds to the majority of individual results. The probability of error in such a scheme scales as $\C[maj] \propto \lvert\braket{\psi_+}{\psi_-}\rvert^N$.

In comparison, the collective Helstrom measurement, performed at once on the collective state of all copies, $\ket{\psi_\pm}^{\otimes N}$, achieves a probability of error $\C[col] = (1-\sqrt{1-\lvert\braket{\psi_+}{\psi_-}\rvert^{2N}})/2$. While this strategy is strictly optimal, with $\C[col]$ scaling as $\lvert\braket{\psi_+}{\psi_-}\rvert^{2N}$ for $N \rightarrow \infty$, it is not in general obvious whether the Helstrom measurement on $N>1$ copies can be performed using only local measurements, making it difficult to practically realize using current experimental techniques.

For this special case of discriminating pure nonorthogonal quantum states, $\ket{\psi_+}$ and $\ket{\psi_-}$, it is possible to achieve the same scaling of error probability as the collective Helstrom measurement by using a fixed local measurement (as in the majority vote scheme but using a different fixed measurement) applied to all $N$ copies~\cite{Acin2005}.  Furthermore, by performing local measurements on each copy adaptively, choosing the measurement bases to minimize the error probability at each stage (thus being ``locally optimal'' adaptive measurements), it is possible to precisely achieve the optimal performance as defined by the collective Helstrom measurement for any $N$~\cite{Acin2005,Brody1996}.

The above picture becomes much more complicated for mixed states. In this general case, the locally optimal adaptive local measurement scheme can exhibit worse performance than the simple majority vote scheme when the number of copies exceeds some threshold~\cite{Higgins2009}. Clearly, neither scheme is the optimal local measurement scheme in the general case. To derive a local measurement scheme that is \emph{globally} optimal, with the least probability of error for all $N$, requires an analysis based on dynamic programming~\cite{Sniedovich2010}. This was done in Ref.~\cite{Higgins2009}, and tested experimentally, for $N \leq 10$.

Here we analyze the performance of a number of multiple-copy two-state discrimination schemes, focussing on the distinction between local optimization and global optimization as it applies to the resulting probability of error of these schemes. We consider two classes of local measurement schemes: those using identical measurement bases for each copy (fixed measurements) and those in which the measurement bases are changed as each measurement result adds new knowledge (adaptive measurements). This gives us four schemes for consideration: locally optimal fixed (LOF) local measurements, globally optimal fixed (GOF) local measurements, locally optimal adaptive (LOA) local measurements, and globally optimal adaptive (GOA) local measurements. For comparison, we also consider the optimal collective measurement (OCM). For each of these schemes we calculate the error probability for states with various levels of depolarizing mixture, for finite $N$. We then determine the large $N$ scaling of the schemes by directly calculating the corresponding Chernoff bound in each case. Doing so, we verify recent theoretical bounds~\cite{Hayashi2009, Calsamiglia2010} placed on the asymptotic scaling of local measurement schemes. In addition, we analyze the relationship between the LOF and GOF local measurement schemes as mixture is varied, and find that the latter has no advantage in the asymptotic limit for all qubit states with mixture greater than 2\%.

\section{Discrimination Schemes}

Consider a pair of pure qubit states passing through a depolarizing channel~\cite{Nielsen2000}. Let $\alpha \leq \pi/2$ be the angle separating the two states in Hilbert space ($\alpha = \pi/2$ thus corresponds to orthogonal states). Without loss of generality we may represent these states as
\begin{equation}
 \hat \rho_\pm = \half [\hat I + (1 - \nu) (\hat Z \cos \alpha \pm \hat X \sin \alpha)]
\end{equation}
where $\hat X$ and $\hat Z$ are the Pauli operators, and $\nu \in [0, 1]$ quantifies the level of mixture applied by the channel. Given a collection of $N$ identical copies of either the $\hat\rho_+$ or the $\hat\rho_-$ state (with equal probability), our task is to then determine, conclusively and with minimal probability of error, which state was given.

The four local discrimination schemes are defined by a sequence of local projective measurements applied to each of the $N$ copies of the unknown state. A local projective measurement performed in the basis $\{\ketbra{\phi}{\phi}, \ketbra{\phi^\perp}{\phi^\perp}\}$, where
\begin{equation}
 \ket\phi \equiv \cos\phi\ket{0} + \sin\phi\ket{1}
\end{equation}
is parametrized by the angle $\phi \in [0, \pi/2)$. (Due to our choice of Hilbert space basis in defining our states $\hat \rho_\pm$, we can restrict measurements to those with real coefficients in this basis.) Each of the local measurement schemes then constitutes a sequence of measurement angles $\{\phi_n\}$, where $1 \leq n \leq N$. Given the state parameters $\nu$ and $c \equiv \cos\alpha$, Bayesian inference can be used to optimally analyze the measurement results and make a final determination of that state, with some probability of error, $\C = \C_N(\nu, c)$.

\subsection{Optimal Single-Copy and Collective Measurement}

\begin{figure}
 \centering \includegraphics{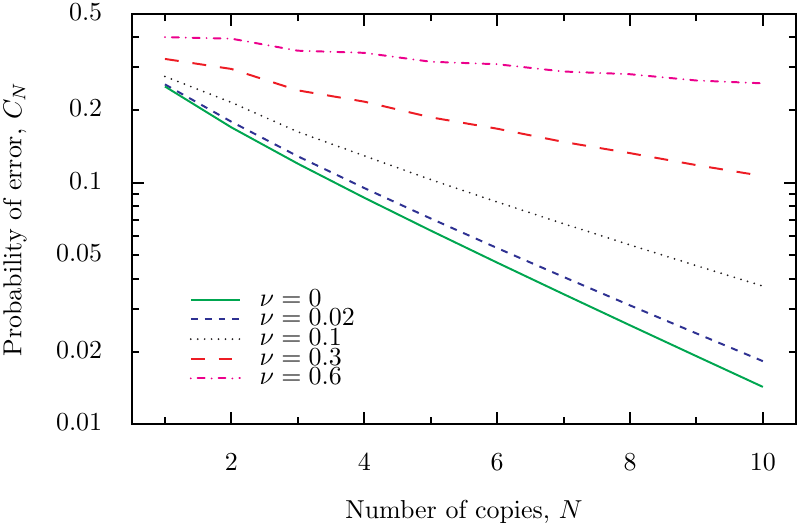}
 \caption{Probability of error of optimal collective measurement of up to 10 copies with $\alpha = \pi/6$ for various amounts of depolarizing mixture $\nu$, and for equally likely states ($q=1/2$).} \label{fig:cocm}
\end{figure}

For any number of copies $N \geq 1$, the minimum possible error probability $\C$ can be obtained in principle by measuring all $N$ copies of the state using a \emph{collective} measurement. A collective measurement is one that, in general, cannot be achieved by local measurements, even allowing for the measurement basis to be chosen adaptively (i.e.\ based on prior results). The optimal collective measurement for any $N$ can be performed by measuring the observable $\hat{\Gamma} = q \hat \rho_+^{\otimes N} - (1 - q) \hat \rho_-^{\otimes N}$ \cite{Helstrom1976,WisMil10} and guessing the state as $\hat\rho_+$ or $\hat\rho_-$ corresponding to the sign of the result obtained. Here $q$ is the prior probability of the systems being in the $\hat\rho_+$ state. The resulting probability of error, known as the Helstrom lower bound, is found to be $\C[OCM]_N = 1 - q + \sum_{j:\gamma_j < 0} \gamma_j$, where $\gamma_j$ are the eigenvalues of $\hat\Gamma$.

Given the conditions $\alpha$ and $\nu$, we can compute the eigenvalues of the operator $\hat{\Gamma}$ and determine the probability of error. Some examples for $\alpha = \pi/6$ and various $\nu$ are shown in Fig.~\ref{fig:cocm}. When $\nu = 0$ (that is, for pure states) the eigenvalues of $\hat\Gamma$ can be determined analytically, and the optimal collective measurement error probability $\C[OCM]_N$ is given by
\begin{equation}
 \left.\C[OCM]_N\right|_{\nu=0} = \half \ro{1 - \sqrt{1 - 4q(1-q)c^{2N}}}, \label{eq:cocm}
\end{equation}
where $c$ (that is, $\cos\alpha$) in this pure state case can also be written $c = \lvert \braket{\psi_+}{\psi_-} \rvert$.

\begin{figure}
 \centering \includegraphics{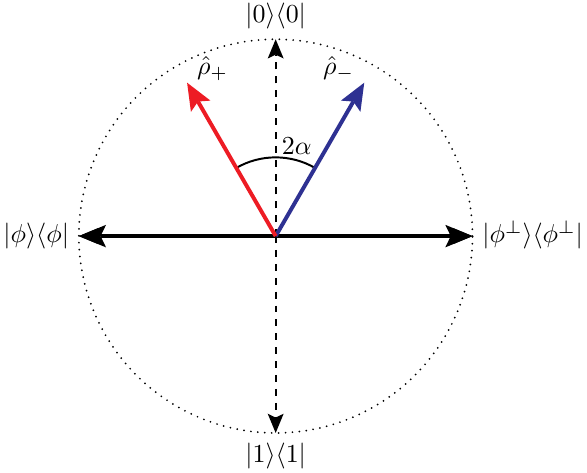}
 \caption{Single-copy state and measurement on a Bloch sphere cross-section. The states $\hat\rho_+$ and $\hat\rho_-$, separated by a Bloch-space angle of $2\alpha$ (corresponding to an angular separation of $\alpha$ in Hilbert space, as used in the text; here $\alpha=\pi/6$), are depicted with $\nu = 0.1$ depolarizing mixture. The measurement basis shown, here with $\phi = \pi/4$, is referred to as an `unbiased' measurement. In the case of equal prior probability ($q = 1/2$), this is the optimal single-copy measurement.} \label{fig:bloch_unbiased}
\end{figure}

\begin{figure}
 \centering \includegraphics{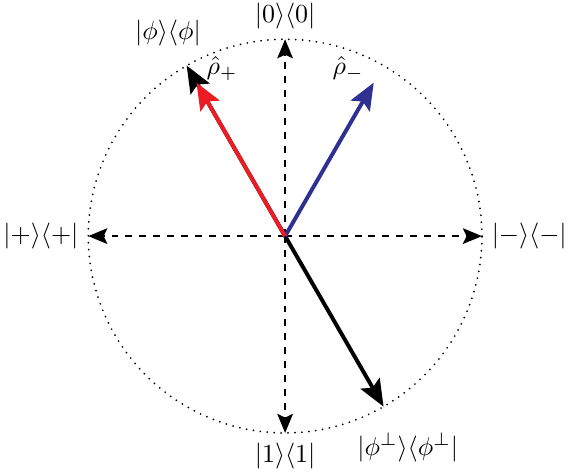}
 \caption{Single-copy state and measurement on a Bloch sphere cross-section, with states $\hat\rho_+$ and $\hat\rho_-$ with $\nu = 0.1$ depolarizing mixture. The measurement basis shown is with $\phi = \alpha/2$, optimal in the limit $q \rightarrow 1$. This and the opposite case where $\phi = \pi/2 - \alpha/2$ (optimal for $q \rightarrow 0$) are referred to as `fully biased' measurements.} \label{fig:bloch_fullybiased}
\end{figure}

The optimal single-copy measurement (OSM) is the special case $N=1$. This is a projective local measurement, independent of $\nu$ due to the invariance of the direction of eigenvectors under depolarization. The measurement angle is given by 
\begin{equation}
 \phi^\text{OSM}(q) = \half\operatorname{arccot}[(2q - 1)\cot \alpha]. \label{eq:phi_osm}
\end{equation}
For the case of equal prior probabilities ($q = 1/2$) the measurement angle $\phi^\text{OSM}$ is $\pi/4$, with $\ketbra{\phi}{\phi}$ and $\ketbra{\phi^\perp}{\phi^\perp}$ symmetric about the state vectors $\hat\rho_\pm$ (see, e.g., Fig.~\ref{fig:bloch_unbiased}). The probability of error is
\begin{equation}
 \C[OSM] = \half \sq{1 - (1 - \nu) \sqrt{1 - c^2}}
\end{equation}
for all $\nu \in [0, 1]$.

We note that in the case of unequal prior probabilities, $q \neq 1/2$, the optimal single-copy measurement angle $\phi^\text{OSM}$ `biases' towards the more likely state. As $q$ increases from $1/2$, the measurement angle decreases such that, in the limit of complete prior determination $q = 1$, $\phi^\text{OSM} = \alpha/2$, with $\ketbra{\phi}{\phi}$ lying in the direction of $\hat\rho_+$ (see Fig.~\ref{fig:bloch_fullybiased}). Similarly, for $q = 0$, $\phi^\text{OSM} = \pi/2 - \alpha/2$, placing $\ketbra{\phi^\perp}{\phi^\perp}$ in the direction of $\hat\rho_-$. We refer to the measurements in these limits as being `fully biased', and the case where $\phi = \pi/4$ as being `unbiased'.

\subsection{Locally Optimal Fixed Local Measurements} \label{sec:flofm}

\begin{figure}
 \centering \includegraphics{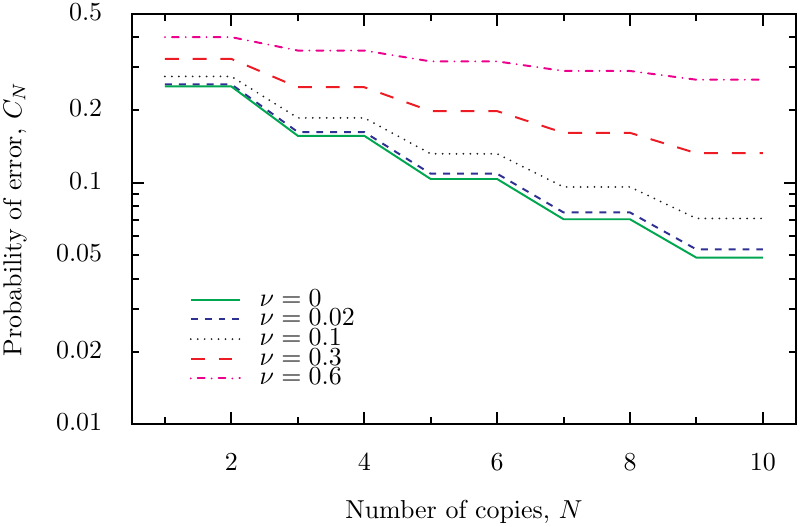}
 \caption{Probability of error of locally optimal fixed local measurements of up to 10 copies with $\alpha = \pi/6$ for various amounts of depolarizing mixture $\nu$.} \label{fig:clofm}
\end{figure}

For any locally optimal scheme, the optimization is done by minimizing the error probability of the measurement of only a single copy. The measurement angles are therefore defined by $\phi^\text{OSM}$ (dependent on $q$). Here we consider the case where each measurement is also performed independently, i.e.\ without any information obtained from the measurement of other copies. Given $\nu$, $\alpha$, and $q=1/2$, all constants, it follows that the measurement bases of such a scheme are fixed for all copies: $\phi^\text{LOF} = \pi/4$.

The LOF scheme reduces to a simple binomial decision problem, with probability of error
\begin{equation}
 \C[LOF]_N = \sum_{n = 0}^{\lfloor N/2 \rfloor} \binom{N}{n} \ro{1-\C[OSM]}^n \ro{\C[OSM]}^{N-n} \label{eq:clofm}
\end{equation}
for $N$ odd. Because the measurements are fixed and symmetric about the states $\hat\rho_\pm$, for $N$ even there is the additional possibility of obtaining measurement results such that Bayesian analysis equally favors both states. In this case, we may arbitrarily choose the overall discrimination result, because the equal a priori likelihood of the states ensures that any choice will lead to the same probability of error. For example, we can randomly choose the outcome. Equivalently, we may choose the state that would have been chosen if the final measurement had not been made, i.e.\ by considering only the first $N-1$ measurement results. In doing so, it is evident that the probability of error for the LOF scheme will not differ from any odd $N-1$ copies to an even $N$ copies, regardless of our method for choosing the overall outcome in this situation. This results in a stepwise decrease of the error probability as the number of copies increases, examples of which can be seen in Fig.~\ref{fig:clofm}, where various levels of depolarizing mixture $\nu$ are applied to the state.

\subsection{Globally Optimal Fixed Local Measurements} \label{sec:fgofm}

To achieve global optimization it is necessary to consider the error probability that arises after all $N$ copies have been measured. One might consider a direct approach, calculating the total error probability by summing over the set of possible outcomes where the Bayesian posterior probability is satisfied (i.e.\ greater than one half, for each outcome), similar to Eq.~(\ref{eq:clofm}), and optimizing over the measurement angle.

Instead of this approach, however, we present an alternative approach based on dynamic programming~\cite{Sniedovich2010}. While more complicated than a direct summation, this approach is considerably more powerful, and will be necessary in following sections. For pedagogical reasons, we introduce it here for the easiest to understand case: a fixed scheme.

The dynamic programming approach allows us to construct a recursive analytic expression for the resulting probability of error. Suppose all $N$ copies are measured in the basis defined by the angle $\phi$, and Bayesian analysis is applied to the first $n \leq N$ measurement results. We obtain a value $P_{n+1}$ that quantifies the credulity that the prepared state is $\hat \rho_+$ at this stage.

Let us define $R_n$ as the probability of error we expect to obtain after measuring the remaining $N-n$ copies. For $n=N$, this is a simple function of the final credulity, $R_N(P_{N+1}) = \min(P_{N+1}, 1-P_{N+1})$. For general $n$, the expected error probability at the $(n-1)$th measurement can be determined from the expected error probability at the $n$th measurement, as
\begin{equation}
 R_{n-1}(P_n) = \sum_{D_n} \Pr[D_n | P_n, \phi] R_n(P_{n+1}), \label{eq:recurs}
\end{equation}
where $D_n$ represents the possible outcomes of the measurement of the $n$th copy and $\Pr[D_n | P_n, \phi] = \Pr[D_{n} | \hat\rho_+, \phi] P_{n} + \Pr[D_{n} | \hat\rho_-, \phi] (1-P_{n})$. In order to evaluate $R_n$, the credulity $P_{n+1}$ can be calculated from the credulity $P_{n}$ and measurement outcome $D_{n}$ by applying Bayes' theorem,
\begin{equation}
 P_{n+1} = \frac{\Pr[D_{n} | \hat\rho_+, \phi] P_{n}}{\Pr[D_n | P_n, \phi]}. \label{eq:bayes}
\end{equation}

Beginning with $n=N$ and progressing backwards towards $n=0$, we can thus calculate the expected probability of error for any $n$ by evaluating Eqs.~(\ref{eq:recurs}) and (\ref{eq:bayes}) at each step using
\begin{equation}
\Pr[D_n | \hat\rho_\pm, \phi] = \frac{\nu}{2} + (1 - \nu) \cos^2 \ro{ \phi - \frac{\pi}{4} (D_n - 1) \mp \frac{\alpha}{2}}.
\end{equation}
In doing so we traverse a binary tree in which all possible permutations of measurement results are visited. The probability of error for $N$ copies of the state is therefore given by
\begin{equation}
 \C_N = R_0(P_1),
\end{equation}
where $P_1=q=1/2$.

\begin{figure}
 \centering \includegraphics{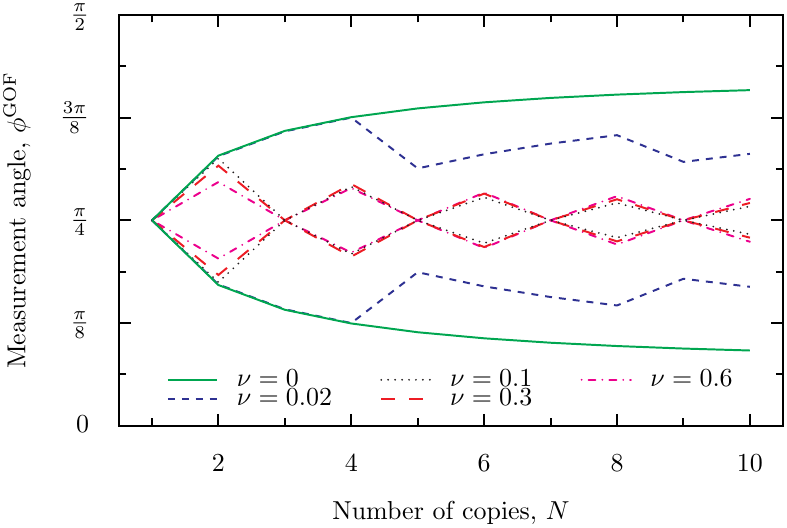}
 \caption{Angles of globally optimal fixed local measurements $\phi^\text{GOF}$ for up to 10 copies with $\alpha = \pi/6$ for various amounts of depolarizing mixture $\nu$.} \label{fig:mgofm}
\end{figure}

\begin{figure}
 \centering \includegraphics{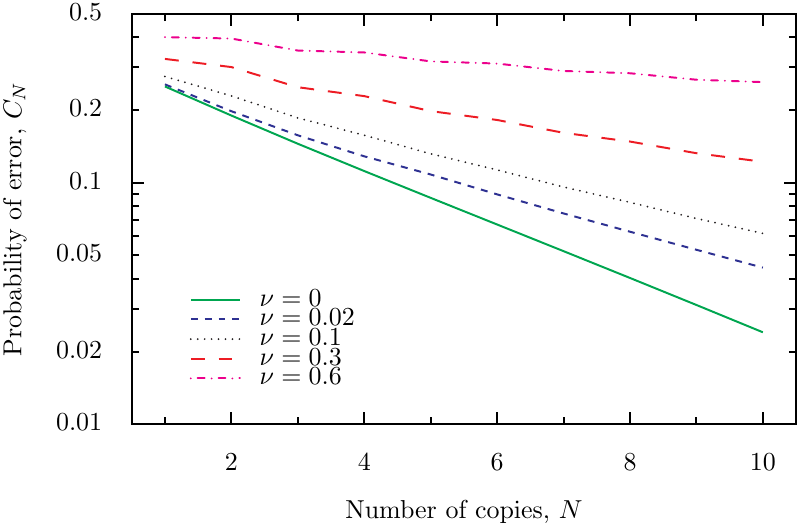}
 \caption{Probability of error of globally optimal fixed local measurements of up to 10 copies with $\alpha = \pi/6$ for various amounts of depolarizing mixture $\nu$.} \label{fig:cgofm}
\end{figure}

Armed with an expression for the probability of error for all $N$ copies, we can determine the globally optimal fixed measurement angle by simply minimizing over $\phi$. In this case, the optimal measurement angle, $\phi^\text{GOF}$, will depend on $N$ and $\nu$, as well as $\alpha$. Figure~\ref{fig:mgofm} shows examples of this globally optimal fixed measurement angle for various amounts of mixture. We see that for $N>1$ in the pure state case, a pair of measurement angles are equally optimal, becoming increasingly biased towards one or the other states as $N$ increases.

For states with appreciable mixture, the measurements oscillate between being unbiased for odd $N$, and somewhat biased for even $N$. In doing so, the scheme avoids the possible condition of obtaining equal credulities for the states following analysis of all measurement results. A mixture of $\nu=0.02$ appears to be an intermediate regime (but see Sec.~\ref{sec:asymp}). Figure~\ref{fig:cgofm} shows the error probabilities for these measurements.

\subsection{Locally Optimal Adaptive Local Measurements}

For the pure state case, $\nu=0$, it has been shown theoretically~\cite{Acin2005} and demonstrated experimentally~\cite{Higgins2009} that, with measurements allowed to vary adaptively, one can achieve an error probability exactly equal to that of the collective measurements for a simple, locally optimal, adaptive scheme. That is,
\begin{equation}
 \left.\C[LOA]_N\right|_{\nu=0} = \left.\C[OCM]_N\right|_{\nu=0} = \half \ro{1 - \sqrt{1 - c^{2N}}}.
\end{equation}
Here each copy is measured in sequence using the locally optimal single-copy measurement given the Bayesian credulity at each stage, $P_n$~\cite{Brody1996}. This measurement basis is defined by the angle $\phi_n^\text{LOA} = \phi^\text{OSM}(P_n)$. The measurement result is used to determine an updated credulity $P_{n+1}$, following Eq.~(\ref{eq:bayes}). This is subsequently used to find the (locally optimal) angle for the next measurement, and so on in turn for each of the $N$ copies.

\begin{figure}
 \centering \includegraphics{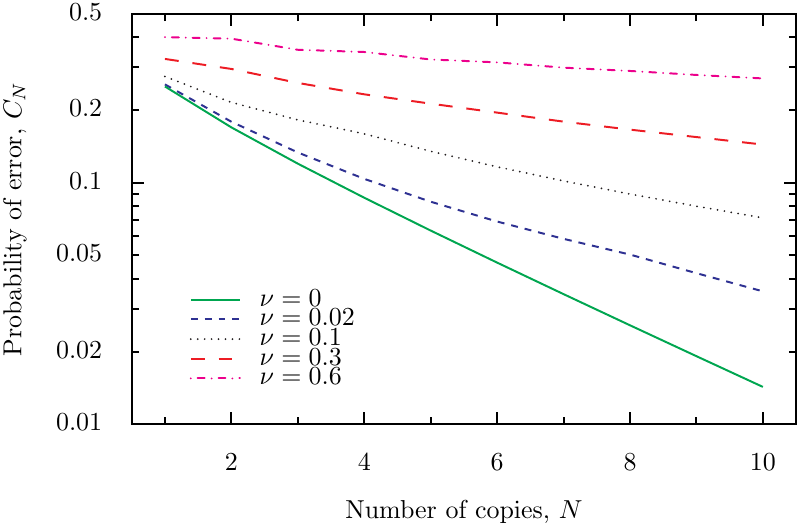}
 \caption{Probability of error of locally optimal adaptive local measurements of up to 10 copies with $\alpha = \pi/6$ for various amounts of depolarizing mixture $\nu$.} \label{fig:cloam}
\end{figure}

For the mixed state case, $\nu > 0$, one could consider a direct approach, in a similar fashion as may be considered for the GOF scheme, by evaluating the error probability over the entire set of measurement outcome strings. Doing so might be feasible for the moderate $N$ considered here, but the complexity of the computation increases exponentially, making such an approach intractable for the large $N$ considered in Sec.~\ref{sec:asymp}.

The dynamic programming approach described in the previous section can also be used to calculate the error probability of this scheme. Here, however, an expression for the measurement angle is already well-defined, given the credulity $P_n$ at the $n$th copy (which can be calculated). The overall probability of error for this scheme can therefore be calculated without requiring a numerical optimization. Examples are illustrated in Fig.~\ref{fig:cloam}.

\begin{figure}
 \centering \includegraphics{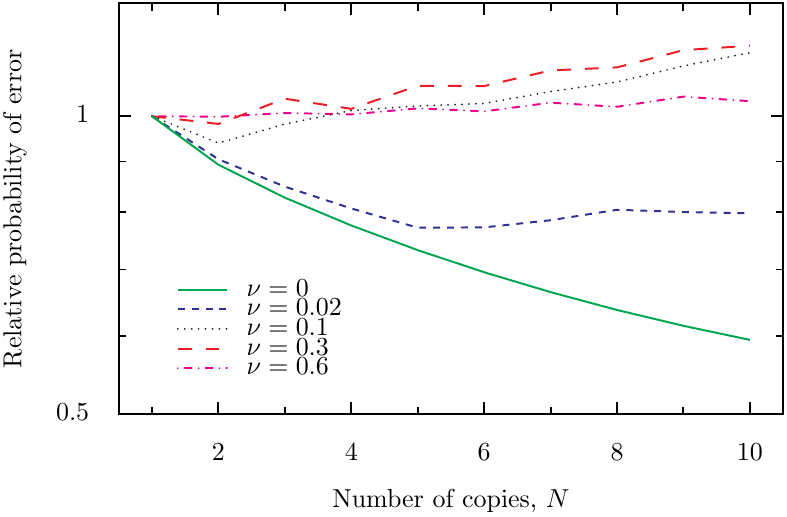}
 \caption{Ratio of the error probability of locally optimal adaptive local measurements and the error probability of globally optimal fixed local measurements, $\C[LOA]_N/\C[GOF]_N$, for various mixtures $\nu$. Values above 1 indicate the LOA scheme exhibiting error probabilities greater (i.e.\ worse) than can be obtained using the GOF scheme.} \label{fig:cloam_rgofm}
\end{figure}

The optimality of the LOA scheme (in terms of lowest overall error probability) does not hold for $\nu > 0$. For certain $\nu$ and $N$ the LOA scheme results in a higher probability of error than a fixed scheme~\cite{Higgins2009}. Figure~\ref{fig:cloam_rgofm} illustrates this by comparing the LOA scheme to the GOF scheme. The error probability for each scheme decreases for each newly measured copy (as each measurement provides information) however there appears a complicated relationship in the ratio of error probabilities of each scheme. The ratio is not monotonic in $\nu$ for $N$ fixed, nor is it monotonic in $N$ for $\nu$ fixed.

\subsection{Globally Optimal Adaptive Local Measurements} \label{sec:fgoam}

To determine the measurements necessary to achieve the lowest error probability possible using local measurements in the general case, we must perform global optimization in a manner similar to that done for the GOF scheme. Relaxing the condition that $\phi$ is fixed, the measurement angle for the $n$th measurement becomes a function of $n$ and the credulity $P_n$, i.e.\ $\phi_n(P_n)$. To consider the potential values of $P_n$ between $0$ and $1$, we construct a table of measurement angles $\phi^\text{GOA}_n(P_n)$ with a large number $s$ of linearly spaced samples of $0 \leq P_n \leq 1$. For intermediate values of $P_n$, the angle $\phi^\text{GOA}_n$ is estimated by cubic interpolation of nearest-neighboring samples.

Bellman's principle of optimality~\cite{Bellman1957,Sniedovich2010} requires that the final stages of an optimal scheme, whatever the conditions are when those stages commence, must necessarily themselves be optimal under those conditions. Adhering to this principle, we construct the table of measurement angles by first recognizing that the final measurement angle $\phi^\text{GOA}_N(P_N) = \phi^\text{OSM}(P_N)$, the optimal single-copy angle, as there is only one copy available (the final copy) to measure at this stage. Thus, entries in the final ($N$th) column of the measurements table represent the optimal single-copy measurement angles $\phi^\text{OSM}(P_N)$ for all $s$ samples of $P_N$.

From this we determine the preceding measurement angle, $\phi^\text{GOA}_{N-1}(P_{N-1})$, by minimizing $R_{N-1}(P_{N-1}, \phi^\text{GOA}_{N-1})$, given $P_{N-1}$. We do this for each $P_{N-1}$ sample, defining the $(N-1)$th column of the measurements table representing $\phi^\text{GOA}_n$. This then defines the optimal penultimate measurement for whatever credulity $P_{N-1}$ may be found by that stage. The $(N-2)$th measurement can be determined similarly by minimizing $R_{N-2}(P_{N-2}, \phi^\text{GOA}_{N-2})$ for all samples of $P_{N-2}$, and so on for the remaining copies. Satisfying the optimality principle at every stage of this reverse construction guarantees that the entire measurement sequence is, of all possible local measurement sequences, the globally optimal one. After constructing the final column of the measurements table, we find the probability of error $\C[GOA]_N = R^\text{GOA}_0(q)$.

It is important to note that the backwards optimizing construction of the measurements table ensures that, for any $n$, the final $n$ measurements are globally optimal. It follows that the values of $R^\text{GOA}_n$ obtained by minimization during this backwards process equal the forwards probabilities of error $\C[GOA]_N$ for each $N=n$. Making use of this fact simplifies the calculation of the measurement angles and error probabilities for multiple  (consecutive) values of $N$. 

\begin{figure}
 \centering \includegraphics{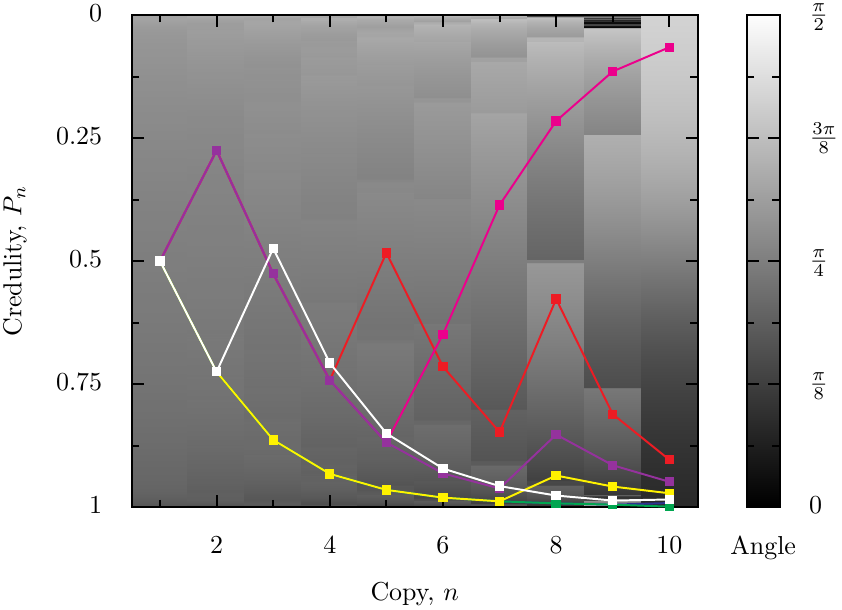}
 \caption{Measurement angles $\phi$ for globally optimal adaptive local measurements of up to 10 copies with $\alpha = \pi/6$ and $\nu=0.1$. As the scheme acquires measurement results it updates the credulity $P_n$, and chooses the subsequent measurement setting corresponding to the shade at the intersection of $P_n$ and $n+1$. The lines illustrate examples of this measurement sequence.} \label{fig:tgoam}
\end{figure}

Figure~\ref{fig:tgoam} illustrates the globally optimal adaptive measurement angles in a mixture regime for up to ten copies, calculated using $s=2501$ samples in the measurement and error probability tables. We found this to be sufficiently accurate as increasing the number of samples made no discernible difference to the outcome for all cases considered here. For large $N$, the initial measurements are approximately unbiased, that is, $\phi \approx \pi/4$, independent of the credulity $P_n$. A complex pattern of measurement angles becomes apparent as more copies are measured, with the final measurement being the optimal single-copy measurement by definition. That is, $\phi(P_n) = \phi^\text{OSM}(P_n)$ of Eq.~\ref{eq:phi_osm}.

Also illustrated in Fig.~\ref{fig:tgoam} are six example trajectories, found by simulating the GOA scheme with measurement results generated randomly according to the outcome probabilities as a function of the state and measurement angles. The actual state is $\hat\rho_+$ in each case plotted. As each copy is measured, the credulity moves in a direction corresponding to the measurement result. In some cases the measurement results are such that the direction is away from the true state. In the average case (and in the asymptotic limit) the credulity $P_n$ increasingly corresponds to the true state.

\begin{figure}
 \centering \includegraphics{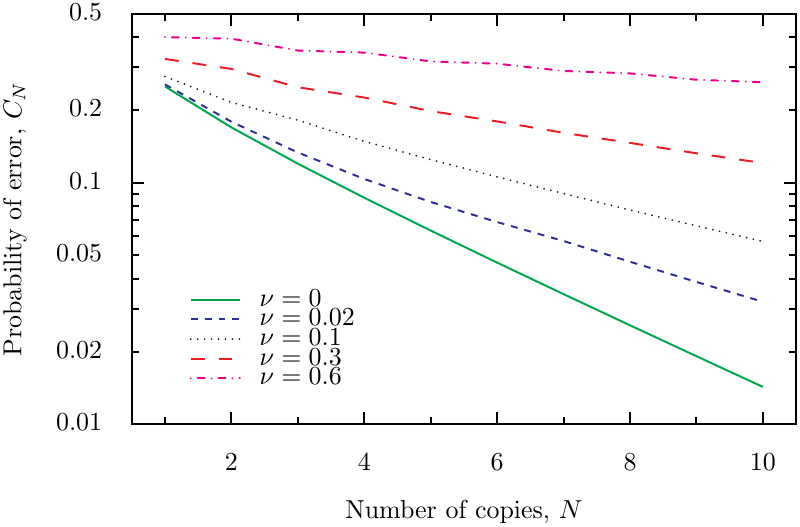}
 \caption{Probability of error of globally optimal adaptive local measurements of up to 10 copies with $\alpha = \pi/6$ for various amounts of depolarizing mixture $\nu$.} \label{fig:cgoam}
\end{figure}

The probability of error for the GOA scheme is plotted, for various levels of mixture, in Fig.~\ref{fig:cgoam}. For pure states this scheme reduces to the LOA scheme, achieving the same error probability as the optimal collective measurement. For mixed states the GOA scheme does not achieve the error probability of the optimal collective measurement, but no other local measurement scheme exhibits lower probability of error.

\section{In The Limit Of Large $N$} \label{sec:asymp}

As the number of copies grows large, i.e.\ $N\rightarrow\infty$, the asymptotic behaviour of the error probability is described by a function decreasing exponentially in $N$~\cite{Chernoff1952},
\begin{equation}
 \C_N \sim e^{-\xi N}.
\end{equation}
The Chernoff bound,
\begin{equation}
 \xi = -\lim_{N\rightarrow\infty} \frac{\log \C_N}{N}, \label{eq:cb}
\end{equation}
corresponding to a given measurement scheme, is a constant quantifying the asymptotic scaling of the error probability, with higher Chernoff bound equating to faster scaling.

Some notable theoretical results already exist in this asymptotic regime. It is known that fixed local measurements are capable of achieving asymptotic scaling equal to that of adaptive local measurements~\cite{Hayashi2009}. This result was recently corroborated by an approach in which the error probability of state discrimination is reformed as a semidefinite programming problem, and for which strict bounds on the error probability were calculated~\cite{Calsamiglia2010} using efficient numerical algorithms~\cite{Vandenberghe1996}. Reference~\cite{Calsamiglia2010} used the same method to investigate other schemes similar to those presented here.

In the following sections we directly calculate the large $N$ behaviour for each scheme considered in the previous sections. We consider optimal schemes under the condition of local and collective measurements given states under depolarizing mixture, and thereby provide a direct verification of the asymptotic results of Refs.~\cite{Hayashi2009} and~\cite{Calsamiglia2010}.

\subsection{Optimal Collective Measurement}

For equally likely states $\hat\rho_+$ and $\hat\rho_-$, the Chernoff bound for optimal collective measurements (i.e.\ the \emph{quantum} Chernoff bound) is~\cite{Audenaert2007}
\begin{equation}
 \xi^\text{OCM} = - \log \min_{0 \leq a \leq 1} \Tr[\hat\rho_+^a \hat\rho_-^{1-a}].
\end{equation}
We can write the states in a diagonal representation, each as a mixture of a pair of orthogonal pure states $\hat\sigma_\pm$ and $\hat\sigma_\pm^\perp$, with $\hat\rho_\pm = (1 - \nu/2)\hat\sigma_\pm + (\nu/2)\hat\sigma_\pm^\perp$. Substituting, we find
\begin{align}
 \xi^\text{OCM} &= - \log \min_{0 \leq a \leq 1} \Tr \Bigl\{ \sq{(1 - \nu/2)^a \hat\sigma_+ + (\nu/2)^a \hat\sigma_+^\perp} \nn
&\quad \times \sq{(1 - \nu/2)^{1-a} \hat\sigma_- + (\nu/2)^{1-a} \hat\sigma_-^\perp} \Bigr\} \\
&= - \log \min_{0 \leq a \leq 1} \Bigl\{ c^2 + (1-c^2) \Bigl[ (1 - \nu/2)^a (\nu/2)^{1-a} \nn
&\quad + (1 - \nu/2)^{1-a} (\nu/2)^a \Bigr] \Bigr\}.
\end{align}

The minimum, which is unique due to convexity~\cite{Audenaert2007}, is satisfied for $a = 1/2$. The Chernoff bound for the optimal collective measurement of mixed states $\hat\rho_\pm$ is therefore
\begin{equation}
 \xi^\text{OCM} = -\log\sq{1-(1-c^2)\ro{1-\sqrt{1-(1-\nu)^2}}}.
\end{equation}
For pure states ($\nu = 0$) this reduces to $-2 \log c$, consistent with the error probability scaling $\C[OCM]_N \sim c^{2N}$ found by taking Eq.~(\ref{eq:cocm}) in the limit of large $N$ (with equal prior probabilities).

\subsection{Locally Optimal Fixed Local Measurements} \label{sec:alofm}

We can rewrite the LOF error probability Eq.~(\ref{eq:clofm}) as
\begin{equation}
 \C[LOF]_N = \ro{\C[OSM]}^N \sum_{n = 0}^{\lfloor N/2 \rfloor} \binom{N}{n} \ro{\frac{1 - \C[OSM]}{\C[OSM]}}^n.
\end{equation}
For large $N$ this has the same scaling as the integral
\begin{equation}
 \C[LOF]_N \sim \ro{\C[OSM]}^N \int_0^{N/2} \binom{N}{n} \ro{\frac{1 - \C[OSM]}{\C[OSM]}}^n \, dn.
\end{equation}
When $n$ is near $N/2$, the binomial coefficient function varies slowly (approximately as a Gaussian of variance $N/4$), but as $n$ approaches $N/2$, $[(1 - \C[OSM])/\C[OSM]]^n$ grows exponentially. The scaling is therefore dominated by this term, and the error probability can be approximated as
\begin{align}
 \C[LOF]_N &\sim \ro{\C[OSM]}^N \binom{N}{N/2} \int_0^{N/2} \ro{\frac{1 - \C[OSM]}{\C[OSM]}}^n \, dn \\
&\sim \ro{\C[OSM]}^N \frac{N!}{[(N/2)!]^2} \frac{[(1 - \C[OSM])/\C[OSM]]^{N/2}}{\log [(1 - \C[OSM])/\C[OSM]]}. \label{eq:clofm_factorials}
\end{align}

To determine the LOF Chernoff bound $\xi^\text{LOF}$ we take the natural logarithm of Eq.~(\ref{eq:clofm_factorials}),
\begin{align}
 \log \C[LOF]_N &\sim (N/2) \log[(1 - \C[OSM])/\C[OSM]] + N \log \C[OSM] \nn
&\quad + \log(N!) - 2\log[(N/2)!] + \text{const.},
\end{align}
where terms constant in $N$ have been omitted. Using Stirling's approximation, this becomes
\begin{align}
 \log \C[LOF]_N &\sim N \log \sq{2\sqrt{(1 - \C[OSM])\C[OSM]}} \nn
&\quad - (1/2) \log N + \text{const.} \label{eq:logclof}
\end{align}
Substituting into Eq.~(\ref{eq:cb}) and taking the limit $N\rightarrow\infty$ we obtain
\begin{align}
 \xi^\text{LOF} &= -\log \sq{2\sqrt{(1 - \C[OSM])\C[OSM]}} \\
&= -(1/2) \log \sq{1 - (1 - \nu)^2 (1 - c^2)}.
\end{align}
For pure states this reduces to $-\log c$ for a probability of error scaling as $\C[LOF]_N \sim c^N$, quadratically worse than $\C[OCM]_N \sim c^{2N}$, consistent with the known scaling~\cite{Higgins2009, Acin2005}.

\subsection{Globally Optimal Fixed Local Measurements}

It has previously been shown~\cite{Calsamiglia2008} that the scaling of the GOF scheme can be quantified by considering the \emph{classical} Chernoff bound~\cite{Chernoff1952} applied to fixed local measurements of each independent copy. Doing so for states $\hat\rho_\pm$ gives
\begin{equation}
 \xi^\text{GOF} = -\log \min_{0 \leq a \leq 1} \, \min_{0 \leq \phi \leq \pi/2} M(a, \phi),
\end{equation}
where
\begin{align}
 M(a, \phi) &= \ro{\Tr[\ketbra{\phi}{\phi}\hat\rho_+]}^a \ro{\Tr[\ketbra{\phi}{\phi}\hat\rho_-]}^{1-a} \nn
&\quad + \ro{\Tr[\ketbra{\phi^\perp}{\phi^\perp}\hat\rho_+]}^a \ro{\Tr[\ketbra{\phi^\perp}{\phi^\perp}\hat\rho_-]}^{1-a}.
\end{align}

\begin{figure}
 \centering \includegraphics{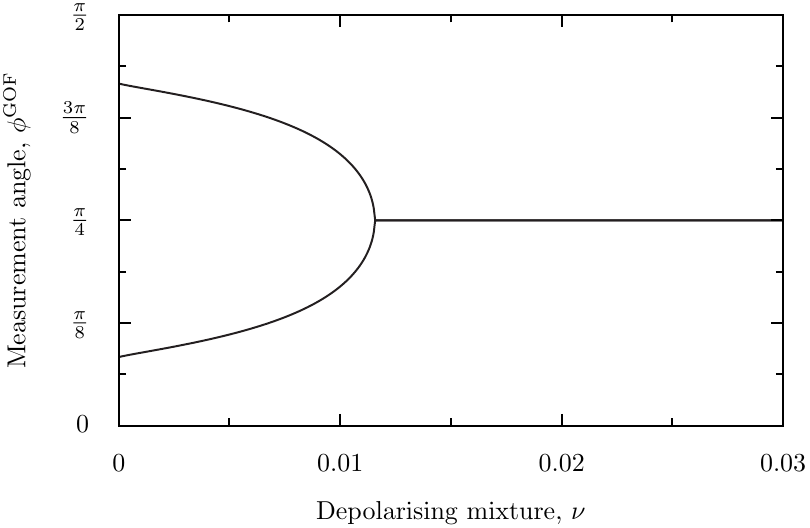}
 \caption{Angles of globally optimal fixed local measurements $\phi^\text{GOF}$ for an unlimited number of copies with $\alpha = \pi/6$ as the amount of depolarizing mixture $\nu$ varies. For pure states there are two equally optimal measurements, fully biased such that in each case one projection is in the direction of one of the states $\hat \rho_\pm$. As mixture increases, the bias of the measurement decreases. At the critical point $\nu^\text{crit}$ the globally optimal measurement angles equal. For any amount of mixture beyond this point the globally optimal fixed measurement angle is $\pi/4$.} \label{fig:magofm}
\end{figure}

We numerically minimize the function $M(a, \phi)$, thus determining $\phi^\text{GOF}$, given the parameters $\nu$ and $\alpha$ that define the states $\hat\rho_\pm$. An example is shown in Fig.~\ref{fig:magofm} for a fixed $\alpha$. For pure states we find that the globally optimal fixed measurement is fully biased---that is, $\phi = \alpha/2$ or $\phi^\perp = \phi - \pi/2 = -\alpha/2$, such that one of the measurement vectors lies in the direction of one of the two states. This is consistent with what we might expect to find given the results plotted in Fig.~\ref{fig:mgofm} for finite $N$. It is also consistent with the proof given in Ref.~\cite{Acin2005} that, for pure states, fully biased measurements exhibit error probability with the same scaling $\xi$ as optimal collective measurements.

For states with sufficient mixture $\nu$, we find the optimal fixed measurement is unbiased, equivalent to the LOF scheme. This confirms the results presented in Ref.~\cite{Higgins2009} that showed unbiased measurements performing better than other schemes, given enough copies of sufficiently mixed states.

\begin{figure}
 \centering \includegraphics{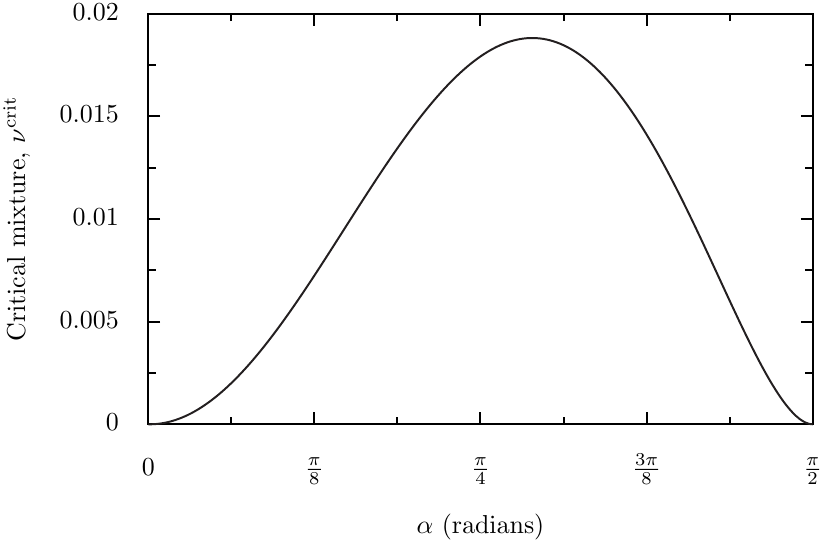}
 \caption{The minimum amount of depolarizing mixture for which the globally optimal fixed local measurement in the asymptotic limit is $\phi^\text{GOF} = \pi/4$ (critical mixture, $\nu^\text{crit}$) as the separation between the states $\alpha$ varies.} \label{fig:critgofm}
\end{figure}

As the purity of the states increases, the optimal fixed measurement angle bifurcates at a critical mixture $\nu^\text{crit}$~\cite{Calsamiglia2010}. For mixture $\nu \ge \nu^\text{crit}$, $\phi^\text{GOF} = \pi/4$. For mixture $\nu < \nu^\text{crit}$ there exists two unique and equally optimal fixed measurement angles related by $\phi^\text{GOF}_1 + \phi^\text{GOF}_2 = \pi/2$. In Fig.~\ref{fig:magofm}, the bifurcation is found at $\nu^\text{crit} \approx 0.012$. As the mixture decreases, the bias of the optimal measurement increases monotonically.

We find that different separations of the states $\alpha$ exhibit different critical mixtures. Figure~\ref{fig:critgofm} shows the dependence of $\nu^\text{crit}$ on $\alpha$, obtained by numeric minimization. The first notable fact is that there is an absolute upper bound on $\nu^\text{crit}$, of about $0.0188$. That is, there is an error threshold of about 2\% depolarizing noise, beyond which there is no advantage asymptotically to using any local strategy beyond the simplest (locally optimal fixed) one, in which the final guess is determined by whichever result occurred in the majority of measurements.

The second notable fact is that $\nu^\text{crit}$ attains this upper bound at an intermediate degree of nonorthogonality of the pure states (prior to depolarization). As the two pure states approach orthogonality ($\alpha = \pi/2$) or identity ($\alpha = 0$), the value of $\nu^\text{crit}$ is zero. That is, any degree of mixture added to such states reduces the optimal solution to a repetition of the optimal single-qubit measurement. In the limit $\alpha\rightarrow\pi/2$, this result is not surprising, as in this limit for pure states the fully biased measurement (which is the globally optimal fixed measurement) becomes the same as the unbiased measurement, because the states become orthogonal. But in the other limit, $\alpha\rightarrow0$, these two measurements are as different as they can be, so the discontinuous change in the optimal measurement as mixture is added is surprising, and may
even be thought paradoxical. What must be borne in mind is that the asymptotic calculation makes no reference to how many copies are required to approach the asymptotic regime. For almost identical states, $\xi^\text{LOF}$ is very small, reflecting the fact that it is very difficult to distinguish the states. From Eq.~(\ref{eq:logclof}) we can see that the asymptotic regime will be reached only when $N/ \log N \gg 1/\xi^\text{LOF}$. That is, as $\alpha\rightarrow 0$, the $N$ required to reach the asymptotic regime diverges. Thus for any fixed $N$, no matter how large, as $\alpha\rightarrow 0$ the problem is necessarily non-asymptotic, and the globally optimal fixed measurement for pure states (the fully biased one) will remain close to the globally optimal fixed measurement for states with a small amount of depolarizing noise.

\subsection{Locally Optimal Adaptive Local Measurements}

\begin{figure}
 \centering \includegraphics{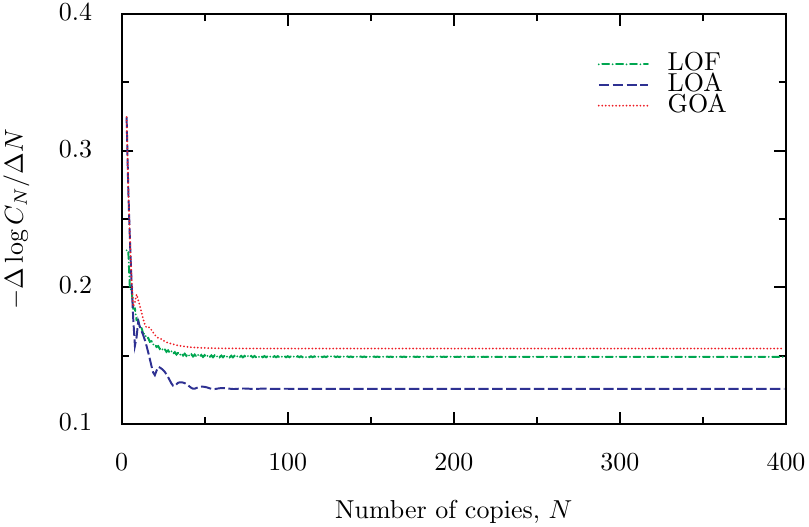}
 \caption{Difference of nearby log error probabilities as a function of $N$, approaching $\xi$ as $N$ grows large. Values are calculated for $\alpha = \pi/6$ and $\nu = 0.02$ using the table sampling and interpolating approximation method with $s=2501$ samples. $\Delta N = 2$ is used to avoid large variations due to the stepwise nature of the LOF error probabilities.} \label{fig:asymp}
\end{figure}

We may find the Chernoff bound $\xi$ for the adaptive schemes by extending the dynamic programming approach (introduced in Sec.~\ref{sec:fgofm}) into the asymptotic regime. We take the difference of the logarithm of error probability between two nearby points in the regime of large $N$, thereby obtaining an estimate of
\begin{equation}
\left.\frac{d \log \C_N}{dN}\right|_{N\rightarrow\infty} = -\xi.
\end{equation}
For the conditions we consider, we find that this gradient reaches a constant (indicating asymptotic behaviour) before $N = 400$ copies---see, for example, Fig.~\ref{fig:asymp}.

For such large $N$, calculation using the recursive approach as detailed in Sec.~\ref{sec:fgofm} becomes computationally infeasible. To avoid this, we instead employ the table sampling and interpolating approximation approach used to determine the error probability of the GOA scheme in Sec.~\ref{sec:fgoam}. For the LOA scheme, the measurement angles $\phi^\text{LOA}_n(P_n)$ are already well-defined, equal to the optimal single-copy measurement angle $\phi^\text{OSM}(P_n)$ where, as before, $P_n$ is the Bayesian credulity of having the $\hat\rho_+$ state given the results of the previous $n-1$ measurements. We may then use this definition to calculate the error probabilities, making use of the fact that $R_n^\text{LOA} = \C[LOA]_{N=n}$. We do so for up to 400 copies.

We have found that, due to the limitations of machine precision, the interpolating approximation becomes problematic when attempting to correctly determine the error probabilities for such large $N$. The problem is best understood by considering the results of this approach as applied to the LOF scheme (i.e.\ with predefined $\phi_n = \phi^\text{LOF}$) as compared to the results we expect due to the exact expression for the error probability. The sharp stepwise pattern of error probabilities, evident in the exact result for this scheme and which can be seen in Fig.~\ref{fig:clofm}, becomes less distinct and ``washed out'' for increasing $N$ when calculating using the interpolating approximation. The outcome is a value for the error probability that is significantly lower than the correct value, and a value for the LOF Chernoff bound, $\xi^\text{LOF}$, significantly greater than the value as determined by our analytic derivation given in Sec.~\ref{sec:alofm}.

We believe that a similar effect also results in the undervaluing of LOA error probabilities and overestimation of the LOA Chernoff bound. This belief is justified as we note that decreasing the number of samples $s$ results in an exaggeration of this effect. Conversely, increasing $s$ reduces the effect, however the reduction appears to be exponentially decreasing in $s$. Indeed, for the LOF scheme, the overestimation is still quite apparent even for $s=10001$ samples. Increasing $s$ quickly becomes computationally infeasible.

To nevertheless extract an accurate estimate of $\xi^\text{LOA}$ from these results, we calculate representative samples, $\xi^\text{LOA}_s$, for various $s \in \{501, 1001, 1501, 2001, 2501, 10001\}$ and fit them to the function
\begin{equation}
 \xi^\text{LOA}_s = x + \frac{y}{(\log s)^{z}}
\end{equation}
using a least-squares method. Here, $x$ and $y$ are allowed to vary. To maintain stability of the solution, $z$ is fixed to the value of $1.22$, which is found to work best.

We finally evaluate the fit function in the limit $s \rightarrow \infty$; that is, we use $\xi^\text{LOA}_\infty = x$. The result of this is a best-guess approximation of the Chernoff bound for the scheme. We have found that this approach works well for all but the highest-purity states, where the projection becomes unstable and the results clearly incorrect (we thus omit those results). The results are shown in Fig.~\ref{fig:chernoff} as a function of $\nu$. As this shows, applying the approach described here to the LOF scheme produces $\xi^\text{LOF}$ that lie only a tiny amount ${\approx}0.0024$ below the exact analytical results, which validates the approach.

\subsection{Globally Optimal Adaptive Local Measurements}

We find the Chernoff bound for the GOA in the same manner as we do for the LOA scheme, by extrapolating fits of the log error probability gradient in the high-$N$ regime. Unlike for the LOA scheme, in this case the measurements are not well-defined in advance, and must be determined by minimizing the error probability. The GOA Chernoff bound is calculated for several numbers of samples $s$, and an extrapolation for $s \rightarrow \infty$ gives us a final estimate of the GOA Chernoff bound, $\xi^\text{GOA}$.

\begin{figure}
 \centering \includegraphics{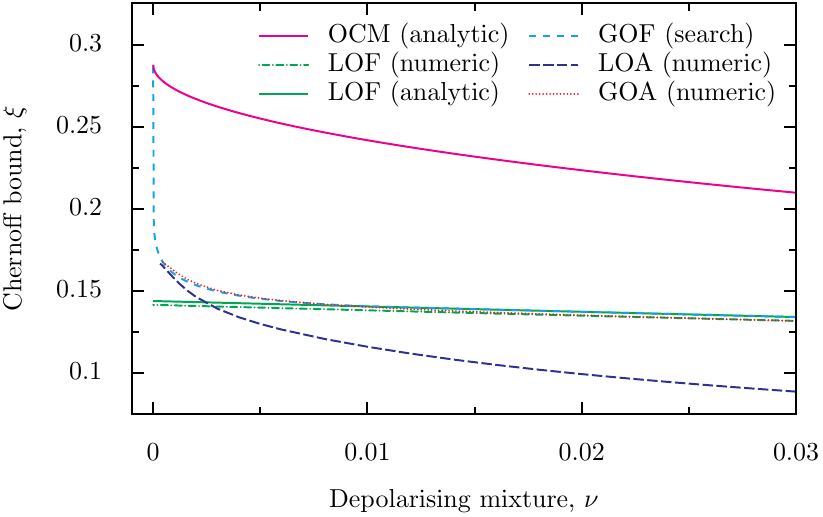}
 \caption{Chernoff bounds for various state discrimination schemes applied to states with $\alpha = \pi/6$ and various levels of mixture. Results for OCM and LOF schemes are calculated analytically, indicated by solid lines (upper and lower, respectively). Results for the LOA and GOA schemes, as well as the LOF scheme for comparison, are calculated using the extrapolated sampling approximation method. Results for the GOF scheme are found by direct optimizing search. The results confirm the GOF and GOA schemes possessing the same scaling in the asymptotic limit.} \label{fig:chernoff}
\end{figure}

The Chernoff bounds for each of the four local schemes are shown in Fig.~\ref{fig:chernoff} for states with $\alpha = \pi/6$. The quantum Chernoff bound of optimal collective measurements is also shown. Optimal collective measurements scale better than any other scheme in all cases except for pure states. For pure states, all but the LOF scheme have a Chernoff bound of $\xi = -2\log c$ (this is $-\log c$ for the LOF scheme). The Chernoff bound of all local measurement schemes rapidly degrades as mixture is introduced. For mixed states, the GOA and GOF schemes have equivalent scaling, optimal for local measurement schemes. For states of high purity, the LOA scheme has better scaling than the LOF scheme, but the converse becomes true as the mixture of the states is increased. For any significant amount of mixture (here greater than approx.\ $\nu=0.008$), the LOF, GOF, and GOA schemes all agree. While the numerically calculated $\xi^\text{GOA}$ is a tiny bit lower than the analytically calculated $\xi^\text{LOF}$ and $\xi^\text{GOF}$, it lies on top of the numerically calculated $\xi^\text{LOF}$, as expected. Thus we have confirmed that, for all but very pure states, the LOF, GOF, and GOA are all optimal local schemes in the asymptotic limit, outperforming the LOA scheme.

\section{Conclusion}

We have investigated the probability of error exhibited by a representative set of local and collective measurement schemes for multiple-copy state discrimination acting on depolarized qubit states. We find that for any nonzero amount of mixture, local schemes fail to exhibit error probabilities approaching that of optimal collective measurements, while the globally optimal adaptive scheme introduced in Ref.~\cite{Higgins2009} outperforms all other local schemes.

We have presented various ways in which we may calculate the error probabilities and measurement settings of the four local measurement schemes for finite $N$. In addition, we have presented analytical expressions and numerical calculations of the asymptotic behaviour (the Chernoff bound) for each scheme. These results provide a direct verification of the asymptotic scaling behaviour predicted of local measurement schemes in works such as Refs.~\cite{Hayashi2009} and~\cite{Calsamiglia2010}. An important conclusion of our work is that there exists a critical degree of mixture, $\nu^\text{crit}_\text{max} \approx 0.0188$, such that, in the asymptotic regime, for any pair of states with at least this much depolarization, the optimal local measurement scheme to distinguish them is the simplest: the locally optimal fixed measurement scheme.

We thank Robin Blume-Kohout and Sarah Croke for helpful discussions. This work was supported by the Australian Research Council, project DP0986503.

\end{document}